\documentclass[prl,aps,preprintnumbers,twocolumn]{revtex4}

\usepackage{amsmath,amssymb,bm}
\usepackage{graphicx}
\usepackage{color}

\begin{document}
\title{Response and phase transition of Kitaev spin liquid in a local magnetic field}
\author{Shuang Liang$^1$}
\author{Bo-Sen He$^1$}
\author{Zhao-Yang Dong$^1$}
\author{Wei Chen$^{1,2}$}

\email{chenweiphy@nju.edu.cn}
\author{Jian-Xin Li$^{1,2}$}
\author{ Qiang-Hua Wang$^{1,2}$}
\email{qhwang@nju.edu.cn}
\affiliation{$^1$Department of Physics, Nanjing University and National Laboratory of Solid State Microstructures, Nanjing, China}
\affiliation{$^2$Collaborative Innovation Center of Advanced Microstructures, Nanjing, China 210093}

\begin{abstract}
We study the response of the Kitaev spin liquid (KSL)  to a local magnetic field perpendicular to the Kitaev honeycomb lattice. 
The local magnetic field induces a dynamical excitation of a flux pair in the spin liquid 
and the system can be described by a generally particle-hole asymmetric interacting resonant level model.
The dynamical excitation of the flux pair closes the flux gap in the spectrum of the spin correlation function locally for the gapless KSL even from the perturbative response to a weak magnetic field. 
 Beyond the perturbative regime, the p-h asymmetry competes with the magnetic field and results in a rich phase diagram. 
 Moreover, the magnetic field breaks the gauge equivalence of the ferromagnetic  and anti-ferromagnetic  Kitaev couplings of the ground state 
 and leads to very different behaviors for the two cases.  
The anti-ferromagnetic case experiences a first order phase transition to the polarized state during magnetization whereas the ferromagnetic case does not. 
 This study can be generalized to the Kitaev model in a uniform magnetic field and may help understand issues in recent experiments on KSL candidates. 
\end{abstract}

\maketitle  
 \section{I. Introduction} 
 
 Quantum spin liquid (QSL) is a strongly correlated system with fascinating properties,
  such as fractionalization,  emergent topological orders, long range entanglement etc.     Since its original proposal as resonating valence bond  liquid state~\cite{Anderson1973}, 
 the search and study of QSL has attracted great efforts from both the theorectical and experimental side~\cite{Zhou2017}. 
 However, due to the lack of any local order, its identification is extremely difficult~\cite{Han2012}. 
 Until about a decade ago, Kitaev proposed an exactly solvable minimal model on a 2D honeycomb lattice, 
 which combines all the features of a QSL yet involves only nearest neighbor interactions on the lattice~\cite{Kitaev2006}.
  This makes it possible to observe the Kitaev spin liquid (KSL) in artifical materials~\cite{Jackeli2009} and cold atom systems~\cite{Duan2003}.

  As one of the characteristic features of quantum spin liquid, the  KSL exhibits fractionalized excitations of gauge fluxes and matter Majorana fermions~\cite{Kitaev2006, Baskaran2007, Knolle2014}. The initiating and probing of such fractionalized excitations often involves the dynamical response of the system~\cite{Wen2017, Ran2017}. 
  The dynamical structure factor of the pure Kitaev model are known exactly and reveals the characteristic fractionalization of the KSL~\cite{Knolle2014}, e.g., a flux gap in the spectrum of the dynamical structure factor appears even for the gapless KSL.  
  However, the dynamics included in the dynamical structure factor of a pure Kitaev model, which is related to an X-ray edge problem by Baskaran et al~\cite{Baskaran2007},  does not involve the dynamics of the flux excitation once the flux is created in the spin liquid. This is not the case for many Kitaev-related models, such as the Kiteav-Heisenberg model~\cite{Jackeli2009, Gohlke2017}, the Kitaev-$\Gamma$ model~\cite{Li2017}.

   In recent experimental search for proximate KSL in real materials~\cite{Wen2017, Ran2017, Yu2018, Ponomaryov2017}, the candidates usually include not only Kitaev couplings but also the above Heisenberg or/and $\Gamma$ interactions~\cite{Li2017, Jackeli2009, Winter2016, Kim2016, Yadav2016, Janssen2017, Brink2016}, which induces a magnetic order at low temperature in the materials. To suppress the magnetic order, the experiments are often conducted in an external magnetic field~\cite{Wen2017, Ran2017, Yu2018, Ponomaryov2017}. In this case, both the extra interaction and the magnetic field has significant impact on the dynamics of the fractionalized excitations.

In this work, we study the KSL in the magnetic field in a simple but non-trivial case, i.e., the KSL in a local magnetic field applied to a single $z$ bond of the honeycomb lattice. We present an analytic study of the response to the magnetic field in such case in both perturbative and non-perturbative point of view, supplemented by exact numerical renormaliztion group (NRG/RG) method. This simple case allows us to have a full understanding of the dynamics and phase transition of the KSL in the local magnetic field and at the same time reveals many generic features of the KSL behavior in a uniform magnetic filed, which may help understand the behavior of the KSL candidates in current experiments.

Our study reveals that the local magnetic field induces interesting physics in the KSL. 
The local magnetic field not only excites a pair of fluxes in the KSL, it also introduces a dynamics to the flux pair, which closes the flux gap in the spectrum of the dynamical spin correlation function locally for a gapless KSL even from the perturbative response to a weak magnetic field. This is in contrast to previous conjectures that such gap is robust against weak magnetic field~\cite{Baskaran2007, Knolle2014}.

The KSL in the local magnetic field is described by a generally p-h asymmetric interacting resonant level model of spinless superconductors. The full dynamics resembles that of a Kondo problem~\cite{Anderson1969, Anderson1970, Anderson1970_2}, but has totally different consequences compared to the conventional Kondo problem in metals~\cite{Anderson1969, Anderson1970, Anderson1970_2, Wiegmann1981}. The p-h asymmetry plays a critical role for the finite $J_z$ case (see models below). It competes with the magnetic field and results in a rich phase diagram in the system. 

The magnetic field  breaks the gauge equivalence of the ferromagnetic (FM) and anti-ferromagnetic (AFM) Kitaev couplings of the KSL ground state and the two cases behave very differently in the local magnetic field. The FM couplings favor the magnetization whereas the AFM Kitaev couplings impede the magnetization. Moreover, the AFM case experiences a first-order phase transition to a polarized state at high magnetic field beyond the perturbation theory as manifested by both the MFT and the NRG results whereas the FM case does not.

This work can be generalized to the  KSL in a uniform magnetic field and may help understand the problems in the current experiments searching for KSL behaviors in real materials. 

 \begin{figure}[t]
\includegraphics[width=3.5cm]{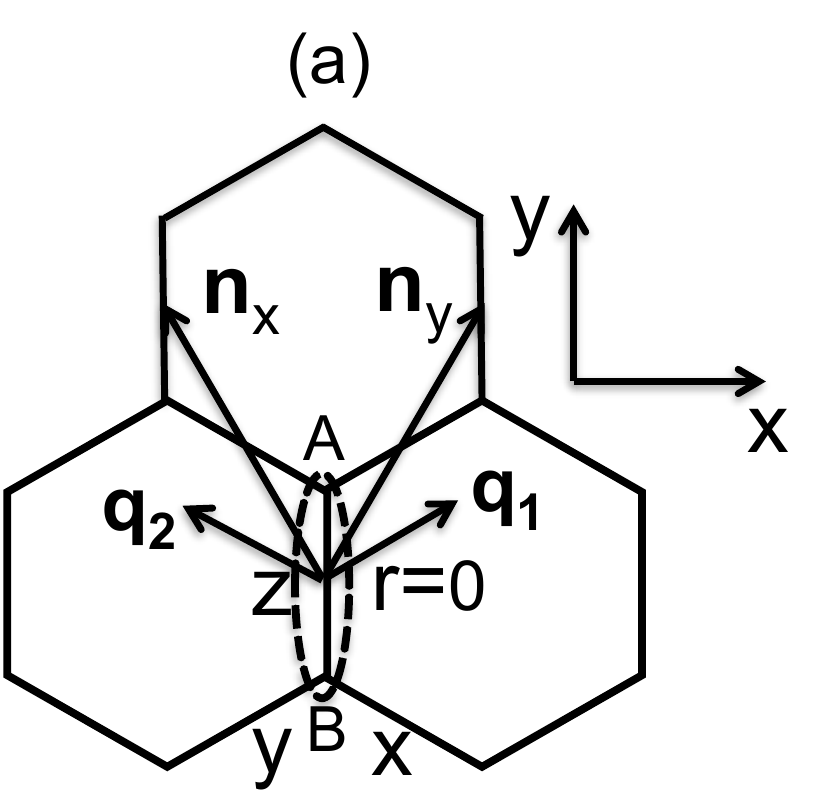}
\includegraphics[width=4.5cm]{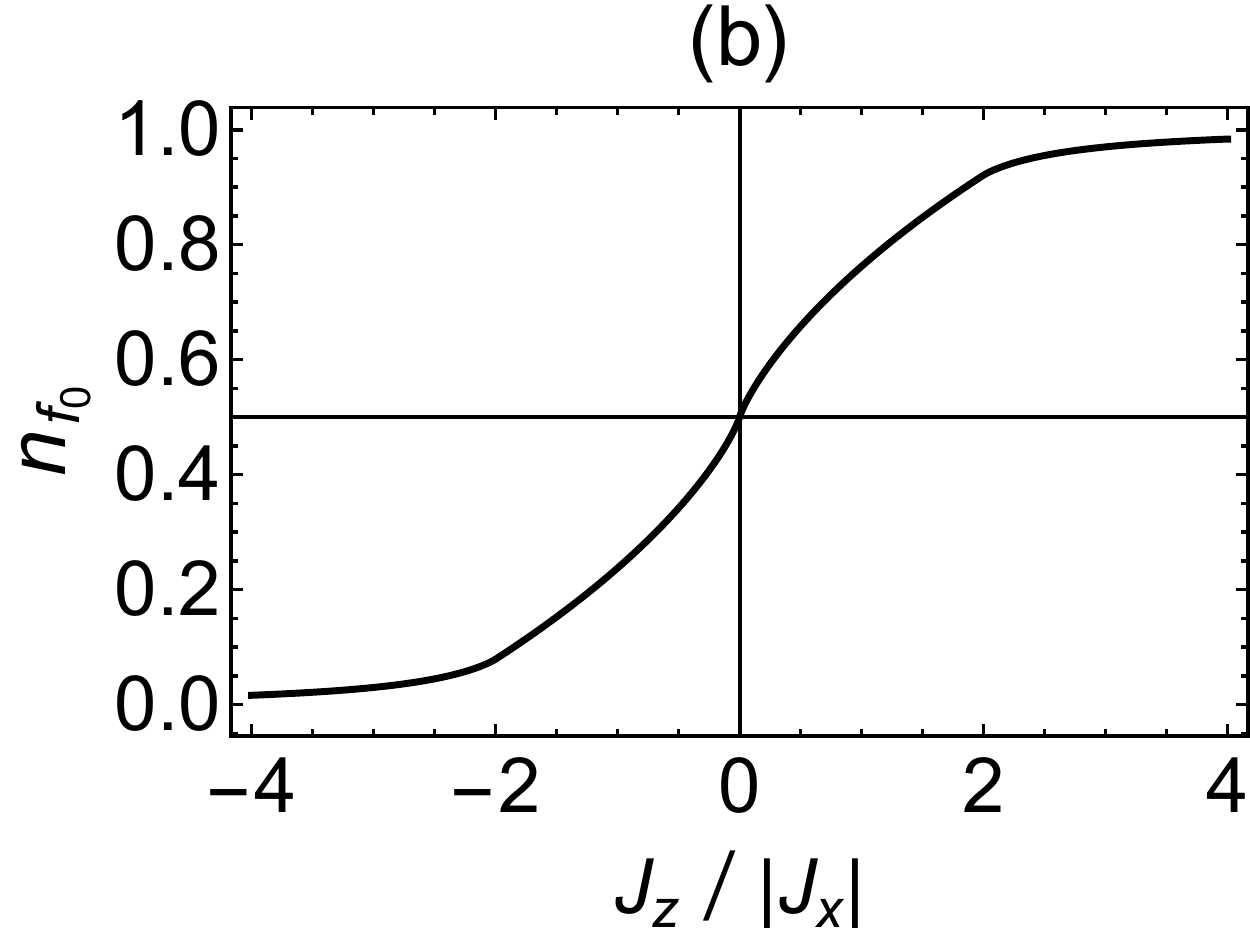}
\caption{ (a)Kitaev honeycomb lattice. The dashed oval represents a unit cell. $\bf{n}_x$ and $\bf{n}_y$ are two unit vectors of the honeycomb lattice and $\bf{q}_1$ and $\bf{q}_2$ are their dual vectors in $\bf{k}$ space. (b) The occupation number of the $f_0$ fermion $n_{f_0}$ as a function of $J_z/|J_x|$ at the Kitaev ground state.
}\label{fig:spectrum}
\end{figure}

\section{II. Model} 
We start by recapitulating the solution of a pure Kitaev model~\cite{Kitaev2006}. The Kitaev Hamiltonian describes a highly  frustrated nearest neighbor interaction of half spins on a honeycomb lattice and takes the form
\begin{equation}
H_K=- \sum_{\langle ij\rangle_\alpha}J_\alpha \hat{\sigma}^\alpha_i\hat{\sigma}^\alpha_j,
\end{equation}
where $\alpha=x,y,z $ denotes three bond directions as shown in Fig.~\ref{fig:spectrum} and $\sigma^\alpha_i$ are the three Pauli matrices on site $i$. $\langle ij\rangle_\alpha$ denotes two sites sharing an $\alpha$ bond.

The model is solved by representing the half spins with four Majorana fermions $\hat{c}_i, \hat{b}^x_i, \hat{b}^y_i, \hat{b}^z_i$ at each site as $\sigma^\alpha_i=ic_i b^\alpha_i$~\cite{Kitaev2006}.   The Kitaev Hamiltonian then becomes
\begin{equation}\label{eq:Kitaev_Hamiltonian}
H_K=i\sum_{\langle ij\rangle_\alpha, \alpha} J_\alpha u_{\langle ij\rangle_\alpha} c_i c_j,
\end{equation}
where the bond operator $u_{\langle ij\rangle_\alpha}\equiv i b^\alpha_i b^\alpha_j$ commutes with the Hamiltonian Eq.(\ref{eq:Kitaev_Hamiltonian}) and is conserved. The ground state entails the gauge invariant quantity $W\equiv\prod_{Plaquette} u_{\langle ij\rangle_\alpha}=1$ for all the plaquettes~\cite{Kitaev2006}. For convenience, we choose the gauge $u_{\langle ij\rangle_\alpha}=-1$ for all the bonds for the ground state in this work. 
There is then no gauge redundancy left in this work.

The Majorana fermions can be combined into two species of complex fermions: bond fermion $\chi_{\langle ij\rangle_\alpha}=i(b^\alpha_i+ i b^\alpha_j)/2$ and matter fermion $f_r=(c_{r,A}+ i c_{r,B})/2$, where $r$ is the unit cell coordinate and $A$ and $B$ are the two sites on the $r$ bond~\cite{Baskaran2007, Knolle2014}.
The ground state Hamilotian has the Bogoliubov de-Gennes (BdG) form when expressed in terms of the matter fermions and describes a spinless p-wave superconductor as
\begin{eqnarray}\label{eq:groundstate}
 H_0&=-&\sum_{r, \alpha=x,y} \left[J_\alpha(f^\dag_r f_{r+n_\alpha} +f_r f_{r+n_\alpha} +h.c.)\right.\nonumber\\
&&\left.\ \ \ \ \ \ \ \ \ \ +J_z(2f^\dag_r f_{r} -1)\right]\nonumber\\
 &=&\sum_{\bf q} \left[\xi_{\bf q} (f^\dag_{\bf q} f_{\bf q}-\frac{1}{2}) +\Delta_{\bf q} f^\dag_{\bf q} f^\dag_{-{\bf q}}+\Delta^*_{\bf q} f_{\bf q} f_{-{\bf q}}\right],\nonumber\\
 \end{eqnarray}
 where $\xi_{\bf q}=-2\rm Re\ \it S_{\bf q}$, $\Delta_{\bf q}=2i\rm Im \ \it S_{\bf q}$ and $S_\bold{q}= J_z+J_x e^{i q_1}+J_y e^{i q_2}$, $q_1, q_2$ are the components of ${\bf q}$ in the ${\bf q}_1$ and ${\bf q}_2$ direction shown in Fig.\ref{fig:spectrum}(a).  
The Hamiltonian Eq.(\ref{eq:groundstate}) can be diagonalized by the Bogoliubov transformation $a_\bold{q} =u_\bold{q} f_\bold{q} -v_\bold{q} f_\bold{-q}^\dag$ with $u_\bold{q} =\cos \theta_\bold{q},\ v_\bold{q} =i \sin \theta_\bold{q},$ and $\tan 2\theta_\bold{q} =\frac{ \rm Im \ \it S_\bold{q}}{ \rm Re \ \it S_\bold{q}}$ ~\cite{Knolle2014},
 \begin{equation}
H_0= - \sum_\bold{q} [|S_\bold{q}| (a_\bold{q}^\dag a_\bold{q} -\frac{1}{2}) -|S_\bold{q}| (  a_\bold{-q}a_\bold{-q}^\dag -\frac{1}{2})].
\end{equation}
The ground state spectrum of the matter fermion is then $\epsilon_0({\bf q})=- |S_\bold{q}|$. The ground state spectrum and energy does not depend on the signs of $J_x, J_y$ and $J_z$ since changing the sign of $J_\alpha$ can be compensated by changing the corresponding $u_{\langle ij\rangle_\alpha}$~\cite{Kitaev2006}. For convenience, we set $J_x=J_y>0$ in this work in the following. The results do not change for $J_{x,y}<0$. For $|J_z|\leq 2 J_x$, the ground state spectrum of the KSL is gapless and for $|J_z| > 2 J_x$, the ground state is gapped~\cite{Kitaev2006}.

We now consider a local magnetic field $\vec{h}$ perpendicular to the honeycomb lattice applied to a single  $z$ bond of the ground state, denoted as $r=0$. The magnetic field Hamiltonian is 
\begin{eqnarray}\label{eq:magnetic_field}
H_h&=&-h(\sigma^z_{0, A}+\sigma^z_{0, B})=-2ih(c_{0, A}b^z_{0, A}+c_{0, B}b^z_{0, B} )\nonumber\\
&=&-2h(f^\dag_0 \chi_{0, z} + \chi^\dag_{0, z}f_0).
\end{eqnarray}
One immediate consequence is that the Hamiltonians with different signs of $J_z$ are now inequivalent.
This is shown as follows: one can transform the Kitaev Hamiltonian with $J_z<0$ to the case $J_z>0$ by the transformation $J_z \to -J_z, b^z_{r,B}\to -b^z_{r,B}$ on all the $z$ bonds. However, the magnetic Hamiltonian then becomes $H_h=-2ih(c_{0, A}b^z_{0, A}-c_{0, B}b^z_{0, B} )=-h(\sigma^z_{0, A}-\sigma^z_{0, B})$, i.e., the magnetic field on the $A$ and $B$ site now has opposite direction. For the reason, the FM case $J_z>0$ and AFM case $J_z<0$ respond very differently to the magnetic field.

 The local magnetic field on the $z$ bond flips the sign of $u_{0,z}=2\chi^\dag_{0,z}\chi_{0,z}-1$ at $r=0$ by creating or annihilating a $\chi_{0, z}$ fermion as shown in Eq.(\ref{eq:magnetic_field}), and thus not only excites a pair of neighboring fluxes $W=-1$ in the ground state but also introduces a dynamics to the flux pair.  For the reason, $u_{0,z}$ is no longer a conserved quantity, whereas $u_{\langle ij\rangle_\alpha}$ on all the other bonds are still conserved. 
We then separate the interaction term on the $z$ bond at $r=0$  from the other terms in the Kitaev Hamiltonian and write the full Hamiltonian as 
\begin{eqnarray}\label{eq:full_hamiltonian}
H&=&H_0+2J_{z} \chi^\dag_{0,z} \chi_{0,z}(2f^\dag_0 f_0 -1)\nonumber\\
&&-2h(f^\dag_0 \chi_{0,z} + \chi^\dag_{0,z}f_0).
\end{eqnarray}
Here $H_0=-i\sum_{\langle ij\rangle_\alpha} J_\alpha c_i c_j$ is the ground state Hamiltonian of the Kitaev model with $u_{\langle ij\rangle_\alpha}=-1$, i.e., $n_{\chi_{\langle ij\rangle_\alpha}}=0$. We have put back the $-iJ_{z} c_{0,A} c_{0,B}$ term to $H_0$ after separating the $-J_z \sigma^z_{0,A}\sigma^z_{0,B}$ term  at $r=0$. The Hamiltonian Eq.(\ref{eq:full_hamiltonian}) describes an interacting resonant level model. The bond fermion $\chi_{0,z}$ acts as an impurity and $\hat{f}_r$ act as itinerant fermions which hybridize and interact with $\chi_{0,z}$ on the site $r=0$. 
The interaction term vanishes for the ground state of the Kitaev model and is non-zero only when there is flux excitation in the system.
For convenience, we denote $\chi_{0, z}$ as $\chi_{0}$ hereafter.

 The eigensectors of the Kitaev Hamiltonian with conserved $u_{\langle ij\rangle_\alpha}$ break the p-h symmetry of the Kitaev model under p-h transformation $\chi_{\langle ij\rangle_\alpha} \to \chi^\dag_{\langle ij\rangle_\alpha}, f_r \to f^\dag_r$ spontaneously. This can be seen from the ground state Hamiltonian Eq.(\ref{eq:groundstate}). The interaction on the $z$ bond reduces to a local potential scattering of $f_r$ fermions in the ground state and breaks the p-h symmetry of the normal state except for $J_z=0$.

The p-h asymmetry of the ground state Hamiltonian $H_0$ is manifested by the occupation number $n_{f_0}$ of the $f_0$ fermions. For the Kitaev ground state, $\langle n_{f_0}\rangle_0=\frac{1}{N}\sum_q |u_q|^2=\frac{1}{2}+ \frac{1}{N}\sum_q \frac{\rm Re \it S_q}{2|S_q|}$ and depends on the ratio $J_z/|J_{x,y}|$ as shown in Fig.~\ref{fig:spectrum}(b). One can see that $\langle n_{f_0}\rangle_0=1/2$ for $J_z=0$ and becomes greater than $1/2$ for $J_z >0$ and less than $1/2$ for $J_z<0$. As $|J_z/J_{x,y}|$ increases, the asymmetry increases.
The deviation of $n_{f_0}$ from half filling  results in an effective positive energy level for the $\chi_0$ fermion that locates in the empty band of matter fermions and thus there is a gap for the flux excitation at $J_z\neq 0$. For the reason, the ground state has $n_{\chi_0}=0$ at $h=0$ and finite $J_z$. 
At the two limiting cases $J_z=\pm \infty$ with $h=0$, the ground state has  $n_{\chi_0}=0$ and $n_{f_0}=\pm1$ respectively corresponding to maximum asymmetry.

At $J_z=0$, the Kitaev model reduces to decoupled one-dimensional chains and its ground state spectrum $\epsilon_0({\bf q})=\pm 2J_x \cos{ \frac{q_x}{2}}$, where $q_x$ is the component of ${\bf q}$ in ${\bf x}$ direction shown in Fig.~\ref{fig:spectrum}(a). The density of states (DOS) of the matter fermion quasiparticles in this case is a finite constant at the Fermi level $E=0$. The resonant level model Eq.(\ref{eq:full_hamiltonian}) is p-h symmetric and can be mapped to the Toulouse limit of a metallic Kondo problem~\cite{Toulouse1969, Wiegmann1981, Giamarchi2004}. At $J_z=0$ and $h=0$, the ground state of Eq.(\ref{eq:full_hamiltonian}) is doublet degenerate corresponding to the $\chi_0$ level empty or occupied. These two states can be considered as the two components of a pseudospin $s_z=1/2-n_{\chi_0}$.
The hybridization of the $\chi_0$ level and $f_0$ fermions by the magnetic field is relevant with scaling dimension $1/2$ at small $h$ and drives the system to the strong coupling (SC) fixed point of a metallic AFM kondo problem at some finite $h$~\cite{Toulouse1969, Anderson1970, Wiegmann1981, Giamarchi2004}. 
The SC fixed point is stable in this case~\cite{Wiegmann1981} and corresponds to a singlet state with the pseudospin $s_z$ fully screened~\cite{Wiegmann1981, Giamarchi2004}.

However, the situation  is very different for  finite $J_z$. 
For $ |J_z|>0$, the Hamiltonian Eq.(\ref{eq:full_hamiltonian}) describes a p-h asymmetric interacting resonant level model of a superconductor.
 Particularly, in the case $0<|J_z|\leq 2J_{x, y}$ the spectrum of the ground state Kitaev Hamiltonian $H_0$ is gapless and
  the DOS of the matter fermions vanishes linearly with energy near the Fermi level~\cite{Kitaev2006}. 
The exact mapping between the resonant level model and Kondo model is lost in this case. 
The dynamics of the former still resembles that of a Kondo problem, but with one species instead of two spin components~\cite{Anderson1969}. 
Such pseudogap Kondo problems had been extensively studied in previous works~\cite{Vojta2004, Ingersent1998, Bulla2001} 
and the p-h (a)symmetry was found to  play a critical role in the system. 

Due to the linearly vanishing DOS of matter fermions, the resonant level model Eq.(\ref{eq:full_hamiltonian}) in the Toulouse limit (i.e., non-interacting limit) has dramatically different consequences in comparison with the case of $J_z=0$. The hybridization between the $\chi_0$ level and the matter fermions is now marginally irrelevant and there is no screening of the $\chi_0$ level in the Toulouse limit in this case~\cite{Vojta2004, Ingersent1998, Bulla2001}. 
Beyond the Toulouse limit, the interaction at finite $J_z $ drives the $\chi_0$ level to positive energy which further 
increases the p-h asymmetry and decreases the effective coupling between the $\chi_0$ and $f_0$ fermion 
thus tends to drive the RG flow near the SC fixed point further away. The SC fixed point at $J_z=0$ and finite $h$ is then unstable 
even under the perturbation of a small $J_z$ in Eq.(\ref{eq:full_hamiltonian}). 

On the other hand, a high magnetic field tends to restore the p-h symmetry because the magnetization is maximum at half filling of $f_0$ and $\chi_0$.
To have a better understanding of the competition between the magnetic field and the asymmetry in the system, 
we study the response of the KSL to the magnetic field in more details in the following.

\section{III. Response to the magentic field} 
The direct response to the magnetic field is the magnetization of the spin liquid. The linear response is related to the dynamic spin-spin correlation function of the bare Kitaev Hamitonian. For the pure Kitaev model, the spin-spin correlation function of an eigensector is nonvanishing only within the bond distance, i.e., $\langle \sigma^\alpha_i(t)\sigma^\beta_j(0)\rangle \sim \delta_{\alpha\beta}\delta_{\langle ij\rangle \alpha}$ ~\cite{Baskaran2007}. This is because $\sigma^\alpha_i$ flips the sign of $u_{\langle ik\rangle_\alpha}$, where $\alpha$ is the bond connecting $i$ and $k$ site. Unless $\sigma^\alpha_i$ and $\sigma^\beta_j$ flip the sign of $u_{\langle ik\rangle_\alpha}$ on the same bond, they create two orthoganal states and  $\langle \sigma^\alpha_i(t)\sigma^\beta_j(0)\rangle$ vanishes.

When a local magnetic field is applied to a single $z$ bond at $r=0$, 
 $u_{\langle ij\rangle_\alpha}$ is conserved for all the bonds except the bond with the local magnetic field. For the reason, the spin correlation function still satisfies $\langle \sigma^\alpha_i(t)\sigma^\beta_j(0)\rangle \sim \delta_{\alpha\beta}\delta_{\langle ij\rangle \alpha}$ and is non-vanishing only within the bond distance. Therefore only the spins on the $z$ bond at $r=0$ are magnetized.

\subsection{1. Perturbative response to the local magnetic field} We first study the perturbative response to the magnetic field.
The linear response to the magnetic field is directly related to the dynamic spin-spin correlation function $g^0_s(t)\equiv\langle 0|T  s^z_{r=0}(t)s^z_{r=0}(0)|0 \rangle$ of the pure Kitaev model where $ s^z_{r=0} \equiv s^z_{0, A}+s^z_{0, B}$.
 This correlation function was studied in previous works in details~\cite{Baskaran2007, Knolle2014}
and it was pointed out by Baskaran et al~\cite{Baskaran2007} that $g^0_s(t)$ can be treated as an X-ray edge problem. A more detailed study of this X-ray edge problem~\cite{Knolle2014} reveals that the spectrum of this spin correlation function exhibits a gap even for the gapless Kitaev spin liquid.
This is clear when one expresses
$g^0_s(t)$ as~\cite{Knolle2014}
\begin{eqnarray}\label{eq:Kitaev_response}
&&g^0_{s}(t)
=-\langle M_0, n_{\chi_{\langle ij\rangle_a}}=0|e^{iH_0 t}(\chi_0f^\dag_0+f_0 \chi^\dag_0)e^{-i H_0 t}\nonumber\\
&&\ \ \ \ \ \ \ \ \ \ \ \ \ (\chi_0f^\dag_0+f_0 \chi^\dag_0)|M_0, n_{\chi_{\langle ij\rangle_a}}=0\rangle \nonumber\\
&&\ \ \ \ \ \ \ \ =-\langle M_0| e^{i E_0 t} f^\dag_0 e^{-i(H_0+V)t} f_0 | M_0\rangle.
\end{eqnarray}
Here $| 0\rangle_0=|M_0, n_{\chi_{\langle ij\rangle_a}}=0\rangle$ is the ground state of the pure Kitaev model 
which includes a matter fermion sector $|M_0\rangle$ and a bond sector $|n_{\chi_{\langle ij\rangle_a}}=0\rangle$.
$E_0$ is the ground state energy of Kitaev model and  $H_0+V=H_0+2i J_z^0 c_{0,A} c_{0,B}$ differs from the ground state Hamiltonian $H_0$
by flipping the bond operator $u_{0,z}$ from $-1$ to $1$. 
This flipping of $u_{0,z}$ creats a pair of neighboring fluxes with $W=-1$ that share the $z$ bond at $r=0$. 
The correlation function $g^0_{s}(t)$ then describes the propagation of matter fermion $\hat{f}_0$ with the presence of a pair of fluxes created at time $t=0$ and annihilated at $t$, and has the same form as the response function of the X-ray edge problem~\cite{Nozieres1, Nozieres2, Nozieres3}.

The spectrum of $g^0_s(t)$ manifests a flux gap even for the gapless Kitaev spin liquid 
when expressed in the Lehmann representation:~\cite{Knolle2014}
\begin{eqnarray}
g^{0}_{s}(\omega)=-\sum_\lambda \langle M_0|f^\dag_0| \lambda \rangle \langle \lambda |f_0 | M_0\rangle
 \delta[\omega -(E_\lambda -E_0)],
\end{eqnarray}
where $| \lambda \rangle$ and $E_\lambda$ are the eigenstates and eigenenergy of the Hamiltonian $H'=H_0+V$ respectively. 
The gap of spectrum function $g^{0}_{s}(\omega)$ is then equal to $\Delta=E_\lambda -E_0$, i.e., the energy difference of  the ground states  of the Kitaev model with and without a flux pair. This gap corresponds to the threshold energy in the X-ray edge problems of metals~\cite{Nozieres1, Nozieres2, Nozieres3}.

A significant difference here from the conventional X-ray edge problems in metals is that  the spectrum of the response function here shows no singularity above the threshold energy (here is the flux gap)~\cite{Knolle2014}. This is because  the linearly vanishing density of states (DOS) of matter fermions near the Fermi level makes the matter fermion correlation function decay as $1/t^2$ on equal site (shown in Appendix) instead of $1/t$ as in ordinary metals~\cite{Gogolin2010, Kitaev2011}. For the reason, the zeroth order of the response function Eq.(\ref{eq:Kitaev_response}) with respect to $V$ behaves as $D^0_{\chi_0}(t)G^0_{f_0}(t)\sim 1/t^2$, where $D^0_{\chi_0}(t)$ and $G^0_{f_0}(t)$ are the equal site Green's function of the $\chi_0$ and $f_0$ fermions with respect to the bare Kitaev Hamiltonian $H_0$ respectively.  The spectrum of this term is then
$\sim(\omega-\Delta)\ln (\omega-\Delta)$ above the threshold energy in contrast to  $\sim \ln (\omega-\omega_{th})$ in a metal~\cite{Gogolin2010}. The higher order contributions of $V$ to the X-ray edge response function can be resummed by the linked cluster theorem to an exponential of the zeroth order contribution~\cite{Gogolin2010}.
For the reason, the spectrum of the response function Eq.(\ref{eq:Kitaev_response}) here results in no singularity above the threshold energy whereas the corresponding X-ray edge response function in a metal results in a power law singularity $\sim (\omega-\Delta)^{-\delta}$, where $\delta$ is related to the scattering phase of the electrons by the hole created by the X-rays and  is positive. The leading order of the spectrum of Eq.(\ref{eq:Kitaev_response}) above the flux gap is $\sim (\omega-\Delta)\ln (\omega-\Delta)$ and increases almost linearly with $(\omega-\Delta)$ above the flux gap for gapless KSL as shown in Ref.\cite{Knolle2014, Knolle2015}.

However, the local magnetic field induces a dynamics to the flux pair which the linear response function does not capture. 
It was shown that the dynamics induced by a uniform magnetic field perpendicular to the honeycomb lattice plane can close the flux gap in the spectrum of the dynamical structure factor of a gapless KSL~\cite{Kitaev2011}. Here we show that this result can be generalized to the KSL in a local magnetic field, i.e.,  a local magnetic field can close the flux gap of a gapless KSL locally.  
This can be seen by expanding the dynamical spin correlation function up to second order of the magnetic field
\begin{equation}\label{eq:second_order_correlation}
g^{(2)}_s(t)=-\frac{h^2}{2}\int\int d\tau_1 d\tau_2 \langle T s^z_0(t) s^z_0(0) s^z_0(\tau_1) s^z_0(\tau_2) \rangle_0.
\end{equation}
The correlation function $g^{(2)}_s(t)$ describes two cycles of X-ray edge absorption and emission processes in succession~\cite{Nozieres1, Anderson1969}.
The asymptotic behavior of $g^{(2)}_s(t)$ at large $t$ can be obtained by generalizing the calculation of Ref.~\cite{Kitaev2011} to here. 
The four operators $s^z_0(\tau)$ in Eq.(\ref{eq:second_order_correlation}) flip the sign of  $u_{0,z}$ at $t=\tau$ with time order respectively and thus create a piecewise-constant potential function of time $V^{(2)}(\tau)$ with and without the flux pair alternatively in the system. 

For a typical time order, e.g. $t>\tau_2>\tau_1>0$, $g^{(2)}_s(t)$ can be expressed as
\begin{widetext}
\begin{equation}\label{eq:perturbative_correction}
g^{(2)}_s(t)=-8h^2\int\int d\tau_1 d\tau_2 
\langle T f^\dag_0(t) f_0(\tau_2) f^\dag_0(\tau_1) f_0(0) e^{-i\int V^{(2)}(\tau) d\tau} \rangle_0, 
\end{equation}
\end{widetext}
where the potential $V^{(2)}(\tau)$ is a piecewise-constant function of time as follows: $V^{(2)}(\tau)=V$ in the interval $(0, \tau_1)$ and $(\tau_2, t)$, and $V^{(2)}(\tau)=0$ at all the other time.
The diagram expansion of the integrand $\tilde{g}^{(2)}_s(t)\equiv \langle T f^\dag_0(t) f_0(\tau_2) f^\dag_0(\tau_1) f_0(0) e^{-i\int V^{(2)}(\tau) d\tau} \rangle_0=e^{C_2} L_2$~\cite{Kitaev2011, Gogolin2010}. Here $e^{C_2}=\langle T e^{-i\int V^{(2)}(\tau) d\tau} \rangle_0$ is the closed loop contribution and $L_2$ is the sum of open line diagrams. The asymptotic form of $e^{C_2}$ is $e^{-i\Delta(\tau_2-t-\tau_1)}$, where $\Delta$ is the energy difference of the ground states with and without the flux pair, i.e., the flux gap. The factor $e^{C_2}$ then raplidly oscillates unless near the points $\tau_1\approx 0$ and $\tau_2\approx t$ within the time interval $\sim 1/\Delta$.

The open line diagram $L_2$ is given by the following equation:
\begin{eqnarray}\label{eq:L_2}
L_2&=&\langle T f^\dag_0(t) f_0(\tau_2) \rangle^{(2)}\langle T f^\dag_0(\tau_1) f_0(0) \rangle^{(2)}\nonumber\\
&&-\langle T f^\dag_0(t) f^\dag_0(\tau_1) \rangle^{(2)}\langle T f_0(\tau_2) f_0(0) \rangle^{(2)}\nonumber\\
&&-\langle T f^\dag_0(t) f_0(0) \rangle^{(2)}\langle T f_0(\tau_2) f^\dag_0(\tau_1) \rangle^{(2)}
\end{eqnarray}
where $\langle T...\rangle^{(2)}$ stands for $\langle T... e^{-i\int V^{(2)}(\tau) d\tau}\rangle e^{-C_2}$.

Due to the oscillation of the $e^{C_2}$ factor, the main contribution to the integration of $g^{(2)}_s(t)$ for large $t$  comes from the regime $\tau_1\approx 0$ and $\tau_2\approx t$, i.e., a small neighborhood of the boundary of $\tau=0$ and $\tau=t$. For such time arguments,
the external potential $V^{(2)}(\tau)$ as a function of $\tau$ turns on only for two short intervals of the order $\Delta^{-1}$
while the seperation between the pulses is large. In this case, the correlator $\langle T...\rangle^{(2)}$ in Eq.(\ref{eq:L_2}) reduces to $\langle T...\rangle_0$ times a trivial renormalization factor $\sim1/J$~\cite{Kitaev2011}.

The bare correlation function of $f_0$ fermion for the typical gapless KSL with $J_z \neq 0$ is obtained in the Appendix as $\langle T  f^\dag_0(t) f_0(0)\rangle_0\sim 1/t^2$  for large $t$ and  $\langle T  f_0(t) f_0(0)\rangle_0=\langle T  f^\dag_0(t) f^\dag_0(0)\rangle_0=0$. We then get the asymptotic form of $g^{(2)}_s(t)\propto A_1 + A_2 \frac{1}{t^4}$ at large t, where $A_1$ and $A_2$ are two constants. The correction to the spectrum of the  spin correlation function due to the irreducible part of $g^{(2)}_s(t)$, i.e., the part subtracting the equal time average, is then $g^{(2)}_s(\omega)\sim \omega^3$ at low energy. We then see that the gap in the spectrum of the dynamical spin-correlation function on the bond with local magnetic field is closed even in a weak local magnetic field. This is in contrast to conjectures in some previous works that this flux gap of the dynamical spin correlation function is robust against weak magnetic field in a gapless KSL~\cite{Baskaran2007, Knolle2014}.

Higher order responses of the Kitaev spin liquid to the local magnetic field contain a succession of X-ray edge absorption and emission processes 
and can be expressed as a determinant of the bare two point correlation function $G^0_{f_0}(\tau_1, \tau_2)$ of the matter fermion $f_0$~\cite{Anderson1969, Anderson1970_2}. The full dynamics of the Kitaev spin liquid in the local magnetic field resembles that of a Kondo problem~\cite{Anderson1969}. However, 
in contrast to the Kondo problem in metals, where this determinant has the Cauchy form and can be computed to all orders~\cite{Anderson1969, Anderson1970_2}, the $G^0_{f_0}(\tau_1, \tau_2)$ for the gapless KSL is proportional to $1/(\tau_1-\tau_2)^2$ and the corresponding determinant is very difficult to evaluate. For the reason, we turn to a different approach to study the response to the magnetic field beyond perturbation regime in the next section.

\subsection{2. Phase transition beyond the perturbative regime}
In this section, we study the magnetization of the Kitaev spin liquid due to the local magnetic field beyond the perturbative regime.

{\it (a) $J_z=0$.}
The $J_z=0$ case reduces to an exactly solvable one dimensional system with zero flux gap.
The total magnetization on the $z$ bond at $r=0$ is 
\begin{eqnarray}
M&=&\langle (\sigma^z_{0, A}+\sigma^z_{0, B})\rangle=\langle 2(f^\dag_0 \chi_{0, z} + \chi^\dag_{0, z}f_0)\rangle\nonumber\\
&=&4\rm \ Re\ \it G_{\chi_0 f^\dag_0}(\tau\to 0^-),
\end{eqnarray}
where $G_{\chi_0 f^\dag_0}(\tau)\equiv-\langle T \chi_0(\tau) f^\dag_0(0)\rangle$. It's straightforward to get $G_{\chi_0 f^\dag_0}(i\omega_n)$ from the equation of motion and 
\begin{equation}
G_{\chi_0 f^\dag_0}(\tau\to 0^-)=\frac{1}{\beta}\sum_n \frac{2h G^0_f(i\omega_n)}{i\omega_n-4h^2 G^0_f(i\omega_n)},
\end{equation}
where $G^0_f(i\omega_n)=\frac{1}{N}\sum_{\bf q}\frac{i\omega_n-2 \rm Re \it S_{\bf q}}{(i\omega_n)^2-(\epsilon_{\bf q})^2}$ and $\beta=1/T$. 
At low energy, the one-dimensional spectrum $\epsilon_{\bf q} \approx J_x (q_x-\frac{\pi}{2})$ and we get $M\sim \frac{h}{J_x}\ln \frac{\Lambda}{h}$ at $T=0$ as shown in the Appendix, where $\Lambda\sim J_x$ is the high energy cutoff. The susceptibility at $h\to 0$ diverges logarithmically as $\xi=\partial M/\partial h\sim -\ln h$  indicating the relevance of the hybridization by the magnetic field at $J_z=0$ and small $h$. 
This divergent response is also due to the zero flux gap at $J_z=0$ which makes the flux excitation and magnetization easy at small $h$.

{\it (b) $J_z\neq 0$.}
For $J_z\neq 0$,
we apply a mean field theory (MFT) analysis to investigate the Hamiltonian Eq.(\ref{eq:full_hamiltonian}) supplemented by NRG method. The latter is regarded as an exact numerical technique. The interaction term in Eq.(\ref{eq:full_hamiltonian}) is decomposed to the  Hartree and Fock part at the mean field level as
\begin{eqnarray}\label{eq:MFT}
&&H^{\rm MF}_{\rm int}=H^{\rm Hart}_{\rm int}+H^{\rm Fock}_{\rm int},\\
&&H^{\rm Hart}_{\rm int}=4J_z n_{\chi_0} f^\dag_0 f_0+4J_z (n_{f_0}-1/2)\chi^\dag_0 \chi_0\nonumber\\
&&\ \ \ \ \ \ \ \ \ \ \  \ -4J_z n_{\chi_0} n_{f_0}, \\
&&H^{\rm Fock}_{\rm int}=- J_z m f^\dag_0\chi_0 - J_z m^* \chi^\dag_0 f_0+ J_z |m|^2/4,  \label{eq:Fock}
\end{eqnarray}
where $n_{f_0}\equiv\langle f^\dag_0 f_0\rangle, n_{\chi_0}\equiv\langle \chi^\dag_0\chi_0 \rangle, m\equiv 4\langle \chi^\dag_0 f_0\rangle$ are the mean fields representing the average values of the occupation number of $f_0$ fermion, $\chi_0$ fermion and the magnetization respectively, and will be determined self-consistently. At $h=0$, the mean fields $m=0$, $n_{\chi_0}=0$, $n_{f_0}>1/2$ for $J_z>0$ and $n_{f_0}<1/2$ for $J_z<0$ at the ground state. The values of $n_{f_0}, n_{\chi_0}, m$ vary with the increase of $h$.
At the mean field level, the $f_0$ and $\chi_0$ fermions obtain a chemical potential of $-4J_z n_{\chi_0}$ and $-4J_z (n_{f_0}-1/2)$ respectively from the Hartree term. The Fock term results in an effective magnetic field $h_{\rm eff}=h+J_z m/2$. Since the magnetization $m$ always has the same direction as the external magnetic field $h$, the FM interaction $J_z>0$ enhances the effective magnetic field and favors the magnetization whereas the AFM interaction $J_z<0$ decreases the effective magnetic field and impedes the magnetization. For the reason, the two cases have dramatically different magnetization processes.

\begin{figure}[ptb]
\includegraphics[width=7.9cm]{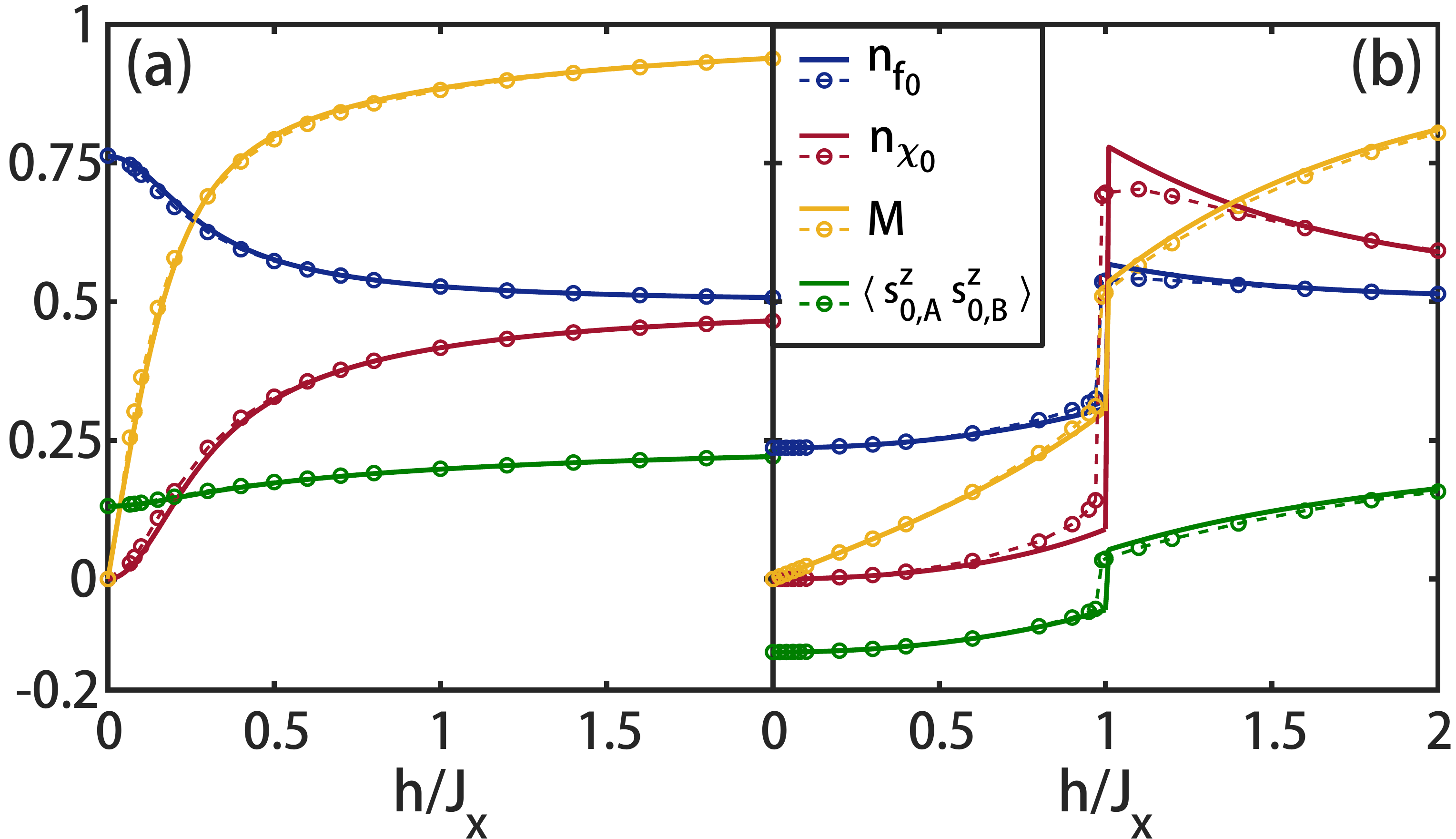}
\caption{(a)$n_{f_0}, n_{\chi_0}, M$ and $\langle s^z_{0,A}s^z_{0,B}\rangle$ as a function of the magnetic field for $J_z/J_x=1$ at $T/J_x= 10^{-4}$. The solid lines are the results from MFT and the dashed lines with circles are from NRG. The unit of $M$ is $2\mu_B$. (b) The same curves for $J_z/J_x=-1$. The legends are the same for the two panels.
}\label{fig:magnetization}
\end{figure}

Fig.~\ref{fig:magnetization} shows the plots of $n_{f_0}, n_{\chi_0}$, $m$ and spin correlation function $\langle s^z_{0,A}s^z_{0,B}\rangle$ as a function of the magnetic field $h$ obtained from the self-consistent MFT for $J_z/J_{x,y}=\pm 1$~\cite{Footnote1}.  
The difference between the FM and AFM coupling case  is clear from the plots. First, the FM case magnetizes much faster than the AFM case as expected since the FM Kitaev coupling enhances the effective magnetic field.
Secondly, the AFM case experiences a first order phase transition at a critical magnetic field $h_{\rm crit}$ with a sharp jump of  $n_{f_0}, n_{\chi_0}$, $m$ and $\langle s^z_{0,A}s^z_{0,B}\rangle$ as shown in Fig.~\ref{fig:magnetization}(b). At the transition both $n_{f_0}$ and $n_{\chi_0}$ jumps from smaller  than half filling to greater than half filling.  The spin correlation function on the $z$ bond with the local magnetic field $\langle s^z_{0,A}s^z_{0,B}\rangle$ jumps from negative to positive value. This jump of $\langle s^z_{0,A}s^z_{0,B}\rangle$ can be detected by the EELS experiments as in Ref.~\cite{Koitzsch2017}.

To check the reliability of the MFT, we  performed NRG calculation, which is exact, for the current problem.
We obtained the exact values of $n_f$, $n_\chi$ and $m$ versus the applied field $h$, as well as the local spin-spin correlation function $\langle s^z_{0,A}s^z_{0,B}\rangle$. The NRG results are shown in Fig.~\ref{fig:magnetization} (dashed lines with circles). 
We see that the results from MFT and NRG agree with each other very well. The possible reason is that the two-particle excitations are gapped in the presence of the applied field, rendering quantum fluctuations beyond the MFT insignificant. However, we find the single-particle excitation spectrum is quantitatively different, even for $h=0$. For example, the $\chi_0$-fermion excitation gap is roughly $J_z$ in MFT, while it is reduced to about $0.26 J_z$ in NRG.

From the mean field Hamiltonian, we can see that the phase transition for the case $J_z<0$ is due to the competition between the magnetic Hamiltonian Eq.(\ref{eq:magnetic_field}) and the Hartree and Fock interaction Eq.(\ref{eq:MFT}). At $h=0$, $n_{f_0}<1/2$ for $J_z<0$. To lower the Hartree and Fock interaction, $n_{\chi_0}=0, m=0$ in the ground state. With the increase of $h$ from zero, $n_{\chi_0}$ and $m$ both increases due to the magnetic Hamiltonian. This increases both the Hartree and Fock interaction for $J_z<0$. To minimize the energy increase of the Hartree term, $n_{f_0}$ increases and becomes closer and closer to $1/2$. At some critical value of $h_{\rm crit}$, this gradual increase of $n_{f_0}$ can no longer compensate the energy increase of the system so $n_{f_0}$ jumps to a value greater than $1/2$. This jump of $n_{f_0}$ makes the Hartree energy negative and favors flux excitation so $n_{\chi_0}$ also jumps from below $1/2$ to above $1/2$ accompanied by a transition from AFM spin correlation to FM spin correlation on the $z$ bond with the magnetic field which favors the magnetization.  As a result, the magnetization also jumps at the transition as shown in Fig.~\ref{fig:magnetization}(b).

\begin{figure}[ptb]
\includegraphics[width=8cm]{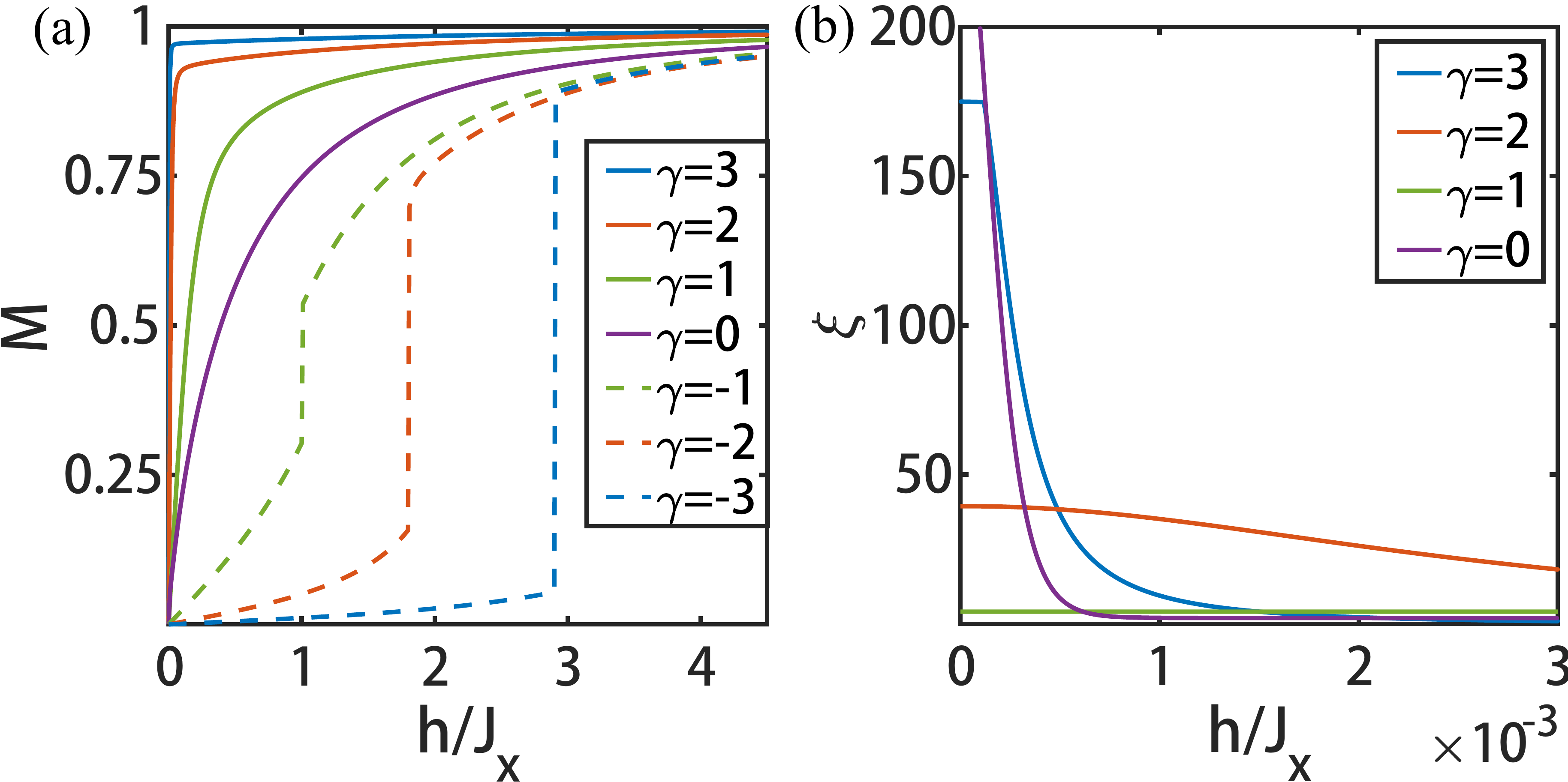}
\caption{(a)The magnetization curves for different $J_z$ from the MFT at $T/J_x=10^{-4}$. $\gamma\equiv J_z/J_x$. (b)The susceptibility curves for $J_z\geq 0$ at small $h$ and $T/J_x=10^{-5}$. The susceptibility diverges for $J_z=0$ and is finite for $J_z>0$ at $h \to 0$. 
 }\label{fig:afm-fm}
\end{figure}

As a comparison, for the FM case $J_z>0$, magnetization increases the Hartree energy yet decreases the Fock energy. The only positive energy is the Hartree energy now. The equilibrium can always be reached by a continuous decrease of $n_{f_0}$ when $h$ increases so there is no phase transition.

We also compared the magnetization curves for different values of $J_z$ as shown in Fig.~\ref{fig:afm-fm}. The $|J_z|\leq 2 J_{x,y}$ cases correspond to gapless spin liquid and the $J_z=\pm 3 J_{x,y}$ cases are gapped spin liquid. We see that at $h\to 0$, the magnetization curve is sharpest for $J_z=0$ with a divergent susceptibility $\xi=\partial M/\partial h$ as expected. However, as $h$ increases, the magnetization curves for $J_z=0$ and $J_z>0$ cross and the $J_z>0$ cases magnetize faster than the $J_z=0$ case. For large $J_z/|J_x|\gg0$, the magnetization saturates very fast at small  $h$. This is because the FM interaction enhances the effective magnetic field by $J_z m/2$ at finite $m$. 
The FM coupling system is then unstable at $J_z\to +\infty$ even under a small magnetic field.

The magnetization curves show no qualitative difference between the gapless and gapped spin liquid for $|J_z|>0$. This is because the p-h asymmetry drives the $\chi_0$ level to the empty band of matter fermions for both cases and the couplings between the $\chi_0$ level and matter fermions have no qualitative difference for the two cases.

We note that at very high magnetic field, both $n_{f_0}$ and $n_{\chi_0}$ tends to half filling for both $J_z>0$ and $J_z<0$ in Fig.~\ref{fig:magnetization}. The p-h symmetry of the Hamiltonian Eq.(\ref{eq:full_hamiltonian}) is then nearly restored at high magnetic field and saturated magnetization.

\section{IV. Discussions and conclusion}
We now have a brief discussion of the RG flow of the Hamiltonian Eq.(\ref{eq:full_hamiltonian}). We note that the parameter $J_z$ enters both the interaction term and the potential scattering of the bare Kitaev ground state Hamiltonian $H_0$ in Eq.(\ref{eq:full_hamiltonian}). The two terms have different scaling dimensions and flow in different ways though they have the same bare coupling $J_z$. At the same time, both terms break the p-h symmetry and generate a finite chemical potential for the $\chi_0$ fermion at finite $J_z$. This chemical potential also flows with scaling. The RG flow of the Hamiltonian Eq.(\ref{eq:full_hamiltonian}) is then described by a high dimensional RG diagram in contrast to the two dimensional RG flow diagrams for usual Kondo problems~\cite{Wiegmann1981, Vojta2004, Ingersent1998, Bulla2001, Vojta2016}. This complicated RG analysis will be left for a future study.

In conclusion, we studied the dynamics of KSL under a local magentic field on a single $z$ bond perperdicular to the Kitaev honeycomb lattice. The local magnetic field leads to a dynamic excitation of a flux pair, which closes the gap locally in the spectrum of the dynamical spin correlation function of a gapless KSL. The system is described by a generally p-h asymmetric interacting resonant level model of spinless superconductors and the dynamics resembles that of a Kondo problem. The p-h asymmetry competes with the magnetic field and results in a rich phase diagram in the system. The magnetic field breaks the gauge equivalence of the FM and AFM Kitaev couplings of the ground state and the two cases behave very differently in the local magnetic field. The ferromagnetic KSL magnetizes much faster in the magnetic field than the anti-ferromagnetic KSL.
The AFM case experiences a first order phase transition during magnetization whereas there is no phase transition for the FM coupling case. 
This study can be generalized to the KSL in a uniform magnetic field and help understand issues in current experiments.

\subsection{V. Acknowledgements.} This work is supported by the NSF of China under Grant No.11504195(WC), No.11774152(JXL), No.11574134(QHW), the National Key Projects for Research and Development of China under Grant No. 2016YFA0300401(JXL and QHW) and Jiangsu Province Educational department under Grant No. 14803002(WC).

\

\begin{widetext}

\section{Appendix}

\subsection{Green's function of the Kitaev model in the ground state}

In this section, we compute the matter fermion Green's function (GF) of the gapless KSL ground state.

The time-ordered GF of matter fermions in the ground state can be calculated as
\begin{eqnarray}
  G_{f}^0(\bold{q}, i \omega_n) &=& -< \hat{T} f_\bold{q}(\tau) f_\bold{q}^\dag(0)>_{i \omega_n},\nonumber\\
                            &=&- < \hat{T} (\cos \theta_\bold{q} a_\bold{q} +i \sin \theta_\bold{q} a_\bold{-q}^\dag  )(\tau) (\cos \theta_\bold{q} a_\bold{q}^\dag -i \sin \theta_\bold{q} a_\bold{-q} )(0)>_{i \omega_n},\nonumber\\
                            &=&\frac{1+ \cos 2\theta_\bold{q}}{2} \frac{1}{i \omega_n +\epsilon_\bold{q}} + \frac{1- \cos 2\theta_\bold{q}}{2} \frac{1}{i \omega_n -\epsilon_\bold{q}},\nonumber\\
                            &=&\frac{i \omega_n - 2 \textrm{Re} \ S_\bold{q}}{(i \omega_n)^2 -\epsilon_\bold{q}^2},
\end{eqnarray}
where $\epsilon_\bold{q}= -2 |S_\bold{q}|$.

The equal site correlation function of $f$ is then
\begin{equation}
  G_{f}^0(i \omega_n; \vec{r},\vec{r}) =\frac{1}{N} \sum_{\bold{q}} \frac{i \omega_n - 2 \textrm{Re} \ S_\bold{q}}{(i \omega_n)^2 -\epsilon_\bold{q}^2}.
\end{equation}

For the gapless KSL with $J_z\neq 0$, the spectrum is linear at low energy and the density of state $\rho(\epsilon) \sim \epsilon$, the GF $ G_{f}^0(Z=i \omega_n; \vec{r},\vec{r})$ at low energy can be easily worked out to be
\begin{equation}
G_{f}^0( Z=i \omega_n; \vec{r},\vec{r})=-\frac{Z}{4\pi \Lambda^2}\left[\ln(\Lambda-Z)-\ln(-Z)+\ln(\Lambda+Z)-\ln Z\right],
\end{equation}
where $\Lambda$ is the high energy cutoff, typically the band width. Analytic continuing to real frequency, we get the time ordered GF
\begin{equation}
G_{f}^0[\omega +i0^+ {\rm sgn}(\omega); \vec{r},\vec{r}]
                        = -\frac{1}{4 \pi \Lambda^2} [2 \omega \ln|\frac{\Lambda}{\omega}| +i \pi \omega \ {\rm sgn}(\omega)].
\end{equation}

The equal site GF in time space for $\tau\gg \tau_0\equiv1/\Lambda$ is obtained by Fourier transformation:

\begin{eqnarray}
G_{f}^0(\tau; \vec{r},\vec{r})  &=&\frac{1}{\beta} \sum_{n}  G_f^0(i \omega_n; 0,0) e^{-i \omega_n \tau},\\
                      &=& \begin{cases} \frac{1}{4 \pi \Lambda^2} ( \frac{\pi}{\beta} )^2 \frac{\cos(\pi \tau/ \beta)}{\sin^2(\pi \tau/ \beta)},\ 0<\tau< \frac{\beta}{2},\\
                                          -\frac{1}{4 \pi \Lambda^2} ( \frac{\pi}{\beta} )^2 \frac{\cos(\pi \tau/ \beta)}{\sin^2(\pi \tau/ \beta)},\ \frac{\beta}{2}<\tau< \beta,
                            \end{cases}
\end{eqnarray}

where $\beta =1 /T$. As $T \rightarrow 0$,  $G_{f}^0(\tau; \vec{r},\vec{r}) \propto 1/ \tau^2$ for large $\tau$.

In  the case $J_z=0$, the DOS of the matter fermion is a finite constant. In this case, the Green's function of the matter fermion is the same as that in ordinary metals and
\begin{eqnarray}
G_{f}^0[Z=i\omega_n; r,r]&=&\frac{1}{\Lambda}\left[\ln(\Lambda+Z)-\ln Z-\ln(\Lambda-Z)+\ln(-Z)\right].
\end{eqnarray}
Analytic continuing to real frequency, one gets $G_{f}^0[\omega+i0^+{rm sgn}(\omega); r,r]=-i\pi \rm Sgn(\omega)/\Lambda$ in this case and $G_{ff^\dag}^0(\tau; \vec{r},\vec{r}) \propto 1/ \tau$ at large $\tau$ at $T\to 0$~\cite{Gogolin2010}.

With the same method, we obtain that the anomalous Green's functions on equal site vanish in the ground state:
\begin{equation}
 F_{f}^0(\tau; \vec{r},\vec{r}) \equiv -< \hat{T} f(\vec{r}, \tau) f(\vec{r},0)>_0=-< \hat{T} f^\dag(\vec{r}, \tau) f^\dag(\vec{r},0)>_0=0.
\end{equation}

\subsection{The magnetic susceptibility at $J_z=0$}

The equal site Green function $G_{ \chi_{0} f_0^\dag}(i \omega_n)$ (we drop the site index $\vec{r}$ in the following) in the case $J_z=0$ is obtained from the equation of motion as:
\begin{equation}
G_{  \chi_{0} f_0^\dag}(i \omega_n)=\frac{2h G_f^0(i \omega_n) }{i\omega_n  -4h^2 G_f^0(i \omega_n)}.
\end{equation}
where $G_f^0(i \omega_n)$ is the equal site Green's function of $f$ of the ground state and has been obtained in the last section.

The equal time correlation function can be obtained by the Fourier transformation:
\begin{equation}
G_{ \chi_{0} f_0^\dag}(\tau\to 0^-) =\frac{1}{\beta} \sum_n G_{  \chi_{0} f_0^\dag}(i \omega_n)
= \frac{1}{\beta} \sum_n \frac{2h G_f^0(i \omega_n) }{i\omega_n  -4h^2 G_f^0(i \omega_n)}.
\end{equation}
Since the above expression has branch cut on the real axis in the complex frequency $Z$ space, we transform  the sum over Matsubara frequency to the integral over two half circles $C_1$ and $C_2$ of the upper and lower half plane of complex frequency space avoiding the real axis of $Z$ as follows:
\begin{align}
\begin{split}
  G_{ \chi_{0} f_0^\dag}(\tau\to 0^-) &=\frac{h}{i\pi }\int_{C_1+C_2} dz \ n_F(z)\frac{G^0_f(z)}{z-4h^2 G^0_f(z)},\\\
                                      &=\frac{2h}{\pi } \int^\infty_{-\infty} d\omega \ n_F(\omega) \textrm{Im} \  [ \frac{G^0_f(\omega+i 0^+)}{\omega+i0^+ -4h^2 G^0_f(\omega+i 0^+)}  ],\\
                                      &=\frac{2h}{\Lambda}\int^\infty_{-\infty} d\omega n_F(\omega)\frac{\omega}{\omega^2+\Gamma^2},
\end{split}
\end{align}
where $n_F(z)$ is the Fermi distribution function and $\Gamma=\frac{4\pi h^2}{\Lambda}, \Lambda=J_x$. At $T\to 0$, $G_{ \chi_{0} f_0^\dag}(\tau\to 0^-) \propto\frac{h}{\Lambda}\ln(\frac{\Lambda^2+\Gamma^2}{\Gamma^2})$.
The magnetization is then $M=4{\rm Re}\ G_{ \chi_{0} f_0^\dag}(\tau\to 0^-) t=\sim \frac{h}{\Lambda}\ln(\frac{\Lambda}{h})$ and the susceptibility $\xi=\partial M/\partial h \propto -\ln h$ as $h\to 0$.

\subsection{Hysteresis loop of the magnetization curves from the MFT for the anti-ferromagnetic Kitaev couplings}

\begin{figure}[b]\label{fig:level_cross}
\includegraphics[width=8cm]{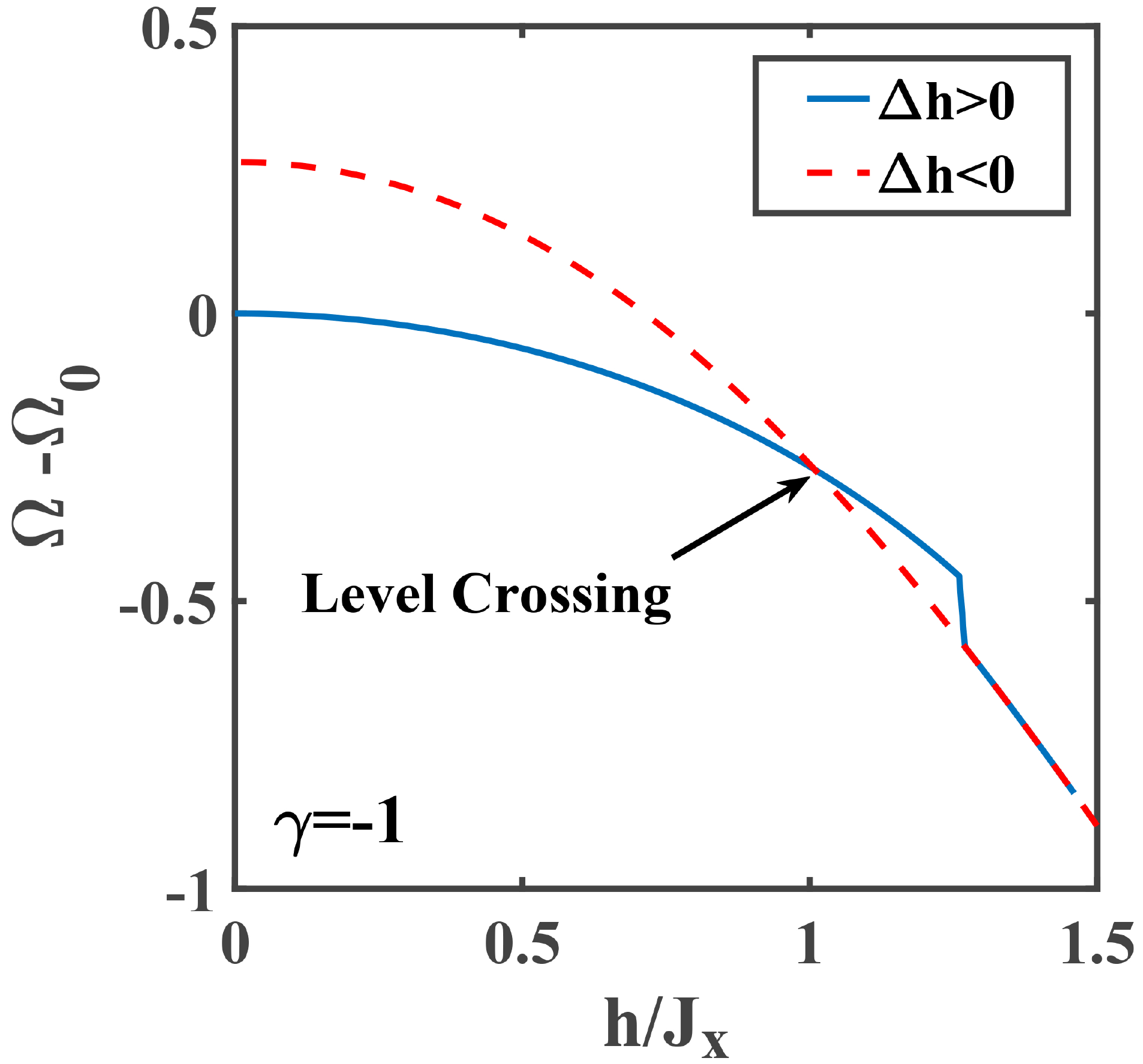}
\caption{The free energy vs. magnetic field $h$ for $h$ going up and down at $\gamma\equiv J_z/|J_x|=-1$. The level crossing point is the global minimum energy point at the transition. At this point, the global minimum energy shows discontinuity from the blue curve to the red dashed curve, which indicates a first-order transition.  
 }
\end{figure}

\begin{figure}[t]\label{fig:hysteresis}
\includegraphics[width=8cm]{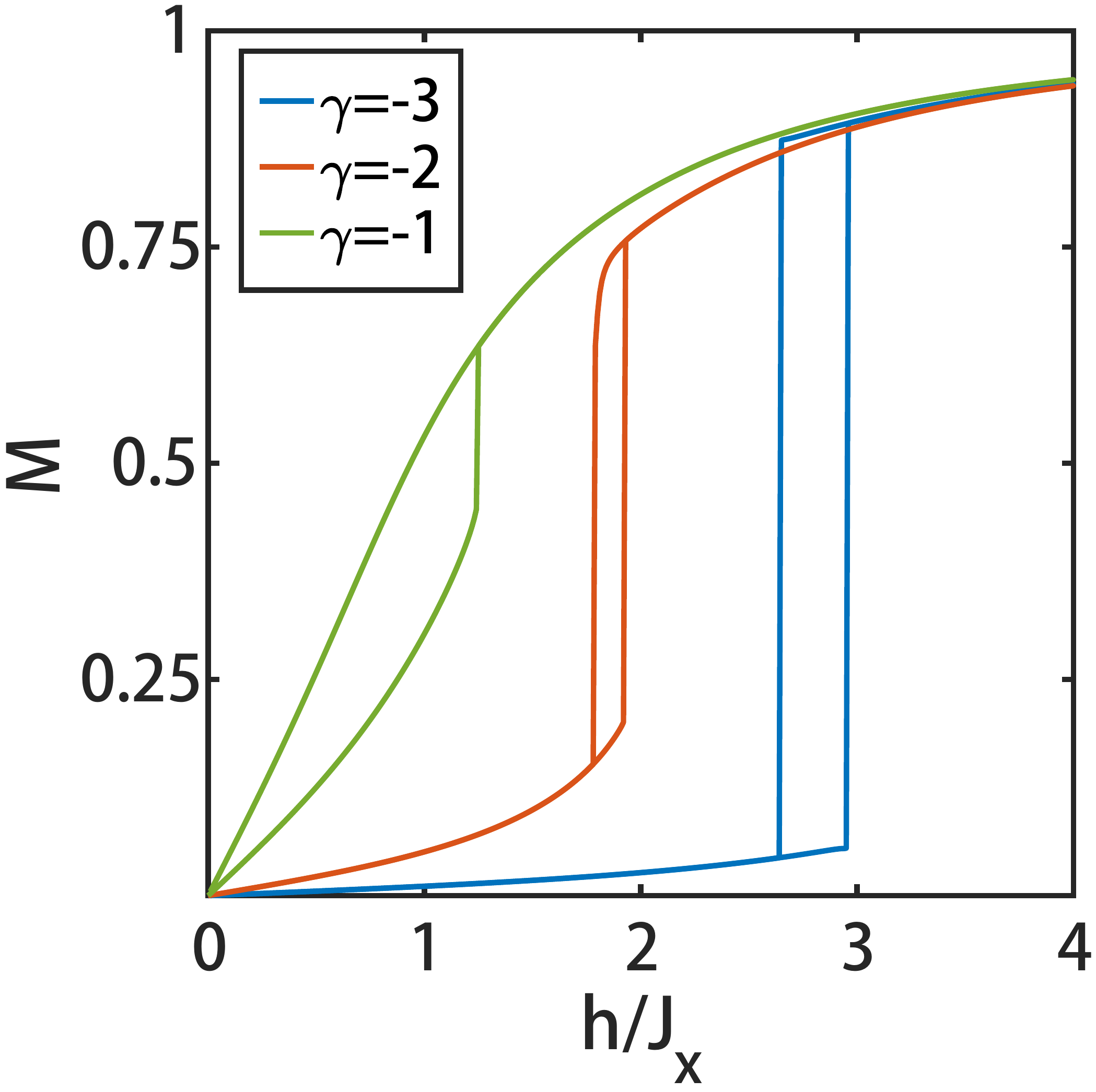}
\caption{The hysteresis loops for the AFM Kitaev couplings $J_z<0$, $\gamma\equiv J_z/|J_x|$. The unit of M is $2\mu_B$.
 }
\end{figure}

In this section, we have a briefly explanation of the magnetization curves obtained from the MFT for the AFM couplings shown in Fig. 2(b) and Fig. 3(a) in the main text. 

The free energy vs. magnetic field shows discontinuity when the magnetic field $h$ is increased or decreased in one direction as shown in Fig. 4. This is typical for a first order phase transition. However, the transition points for the up and down processes are  different due to the superheating and supercooling effects. As a result, the magnetization curves show hysteresis for $h$ going up and down as shown in Fig. 5. The reason for the superheating and supercooling effects is because the free energy in the MFT is trapped in a local minimum, or metastable state before the transition in the calculation. The crossing point of the up and down cycle corresponds to the global minimum of the free energy at the transition and is then the intrinsic MFT phase transition point we presented in the magnetization curves in Fig.2b and Fig.3a in the main text.

\end{widetext}


\begin{thebibliography}{99}
\bibitem{Footnote1}The magnetization shows hysteresis as a characteristic for a first order phase transition due to the superheating or supercooling effect. The transition point shown in Fig. 2 is the intrinsic one from the global minumum of the mean field energy as presented in the appendix.
\end{thebibliography}
\end{document}